\documentclass[preprint,showpacs,preprintnumbers,amsmath,amssymb]{revtex4}
\usepackage{epsfig}
\usepackage{graphicx}
\usepackage{dcolumn}
\usepackage{bm}

\def\beq{\begin{equation}}
\def\eeq{\end{equation}}
\def\eeqn{\end{equation}}
\newcommand\iden{\leavevmode\hbox{\small1\normalsize\kern-.33em1}}

\newcommand{\bea} {\begin{eqnarray}}
\newcommand{\eea} {\end{eqnarray}}

\newcommand{\SLASH}[2]{\makebox[#2ex][l]{$#1$}/}
\newcommand{\Eslash}{\SLASH{E}{.2}}
\newcommand{\sinb}{\sin\beta}
\newcommand{\cosb}{\cos\beta}
\newcommand{\tb}{\tan\beta}
\newcommand{\ctb}{\cot\beta}

\let\jnfont=\rm
\def\NPB#1,{{\jnfont Nucl.\ Phys.\ B }{\bf #1},}
\def\PLB#1,{{\jnfont Phys.\ Lett.\ B }{\bf #1},}
\def\EPJC#1,{{\jnfont Eur.\ Phys.\ Jour.\ C }{\bf #1},}
\def\PRD#1,{{\jnfont Phys.\ Rev.\ D }{\bf #1},}
\def\PRL#1,{{\jnfont Phys.\ Rev.\ Lett.\ }{\bf #1},}
\def\MPLA#1,{{\jnfont Mod.\ Phys.\ Lett.\ A }{\bf #1},}
\def\JPG#1,{{\jnfont J.\ Phys.\ G }{\bf #1},}
\def\CTP#1,{{\jnfont Commun.\ Theor.\ Phys.\ }{\bf #1},}
\def\JHEP#1,{{\jnfont JHEP \ }{\bf #1},}
\def\NPPS#1,{{\jnfont Nucl.\ Phys.\ Proc.\ Suppl.\ }{\bf #1},}
\def\CPC#1,{{\jnfont Comput.\ Phys.\ Commun.\ }{\bf #1},}
\def\CPL#1,{{\jnfont Chin.\ Phys.\ Lett. }{\bf #1},}
\def\APPB#1,{{\jnfont Acta\ Phys.\ Polon.\ B }{\bf #1},}
\def\PR#1,{{\jnfont Phys.\ Rept. }{\bf #1},}

\def\lsim{\raise0.3ex\hbox{$<$\kern-0.75em\raise-1.1ex\hbox{$\sim$}}}
\def\gsim{\raise0.3ex\hbox{$>$\kern-0.75em\raise-1.1ex\hbox{$\sim$}}}
\def\JCAP#1,{{\jnfont JCAP \ }{\bf #1},}

\begin{document}

\title{\ \\[10mm] A simplified 2HDM with a scalar dark matter and
 the galactic center gamma-ray excess}

\author{Lei Wang, Xiao-Fang Han}

\affiliation{
Department of Physics, Yantai University, Yantai 264005, PR China
\vspace{0.5cm} }


\begin{abstract}
Due to the strong constraint from the LUX experiment, the scalar
portal dark matter can not generally explain a gamma-ray excess in
the galactic center by the annihilation of dark matter into
$b\bar{b}$. With the motivation of eliminating the tension, we add a
scalar dark matter to the aligned two-Higgs-doublet model, and focus
on a simplified scenario, which has two main characteristics: (i)
The heavy CP-even Higgs is the discovered 125 GeV Higgs boson, which
has the same couplings to the gauge bosons and fermions as the SM
Higgs. (ii) Only the light CP-even Higgs mediates the dark matter
interactions with SM particles, which has no couplings to $WW$ and
$ZZ$, but the independent couplings to the up-type quarks, down-type
quarks and charged leptons. We find that the tension between
$<\sigma v>_{SS\to b\bar{b}}$ and the constraint from LUX induced by
the scalar portal dark matter can go away for the isospin-violating
dark matter-nucleon coupling with $-1.0< f^n/f^p<0.7$, and the
constraints from the Higgs search experiments and the relic density
of Planck are also satisfied.
\end{abstract}

\maketitle

\section{Introduction}
Over the past several years, a gamma-ray excess at GeV energies
around the galactic center has been identified in the Fermi-LAT data
by several groups \cite{10gev}. The recent study shows that the
excess seems to be remarkably well described by an expected signal
from $31-40$ GeV dark matter (DM) annihilating dominantly to
$b\bar{b}$ with a cross section $<\sigma v>_{b\bar{b}}\simeq
1.7-2.3\times10^{-26}cm^3/s$ \cite{hoopbb}, which is strikingly
close to the thermal relic density value, $<\sigma v> \sim 10^{-26}
cm^3/s$. Since the Higgs couplings to the fermions tend to be
proportional to their masses, the Higgs portal DM is a simple
scenario for DM model which explains the gamma-ray excess. However,
to obtain such large $<\sigma v>_{b\bar{b}}$, the model with the
scalar portal DM will lead to a spin-independent cross section
between DM and nucleon which is excluded by the LUX experiment
\cite{lux}. Therefore, Ref. \cite{1404.3716} consider the
pseudoscalar mediator instead of a scalar, and Ref. \cite{flavorb}
assume that the DM preferentially couples to b-quark. The
measurements of the Higgs invisible width are quite imprecise, and
the invisible branching fraction is required to be smaller than 0.55
from the CMS search for invisible decays of Higgs bosons in the
vector boson fusion and associated $ZH$ production modes
\cite{1408.2106-69}. However, from the analysis of the global fit to
the Higgs signals strengths, the invisible branching fraction is
required to be small than 0.1 at 68\% C.L., see \cite{1408.2106-56}
and \cite{1408.2106}. Therefore, it is challenging for the 125 GeV
Higgs as the mediator since the large Higgs decay into DM is
disfavored by the global fit to LHC Higgs signals.
 The excess of gamma-ray can be also fit by the 10 GeV DM
annihilating to $\tau\bar{\tau}$ \cite{gamtautau}. The various DM
models have been proposed to explain the excess of gamma-ray
\cite{1404.3716,flavorb,onshell,gambbother}.

In this paper, with the motivation of eliminating the tension
between $<\sigma v>_{b\bar{b}}$ and the LUX experiment induced by
the scalar portal dark matter, we add a scalar DM $(S)$ to the
aligned two-Higgs-doublet model (2HDM) \cite{a2h-1,a2h-2}, and focus
on a simplified scenario. Different from the SUSY models, the five
Higgs masses in the 2HDM are theoretically independent. We assume
that the pseudoscalar and charged Higgs are very heavy, and the
heavy CP-even Higgs is the discovered 125 GeV Higgs boson
\cite{125higgs}. The mixing angle $\alpha$ equals to $\beta$, which
leads to that the heavy CP-even Higgs has the same coupling to the
gauge bosons and fermions as the SM Higgs. In addition, we assume
that only the light CP-even Higgs mediates the DM interactions with
SM particles, which has no couplings to $WW$ and $ZZ$, but the
independent couplings to the up-type quarks, down-type quarks and
charged leptons. We show that the tension between $<\sigma v>_{SS\to
b\bar{b}}$ and the constraints from the LUX induced by the scalar
portal DM can go away for the isospin-violating $S$-nucleon
coupling, and the constraints from the relic density of Planck, the
Higgs search at the collider and the other relevant experiments are
also satisfied. Note that Refs. \cite{ii2hdm-dm,iii2hdm-dm} study
the constraint on the Type-II and Type-III 2HDMs with a scalar DM
from the direct detection experiments. In Ref. \cite{htm-dm}, a
scalar DM is added to the Higgs triple model, which gives a valid
explanation for 130 gamma-ray line signal and the enhancement of LHC
diphoton Higgs signal \cite{130ray}.

Our work is organized as follows. In Sec. II we recapitulate the
simplified aligned 2HDM with a scalar DM (S2HDM+D), and analyze the
constraints from relevant experimental constraints. In Sec. III we
give the numerical results, and show that the scalar portal DM in
our model can explain the gamma-ray excess. Finally, we give our
conclusion in Sec. IV.

\section{Simplified two-Higgs-doublet model with a scalar dark matter and
the relevant experimental constraints}
\subsection{Model}
The general Higgs potential is written as [19]
\begin{eqnarray} \label{V2HDM} \mathrm{V} &=& m_{11}^2
(\Phi_1^{\dagger} \Phi_1) + m_{22}^2 (\Phi_2^{\dagger}
\Phi_2) - \left[m_{12}^2 (\Phi_1^{\dagger} \Phi_2 + \rm h.c.)\right]\nonumber \\
&&+ \frac{\lambda_1}{2}  (\Phi_1^{\dagger} \Phi_1)^2 +
\frac{\lambda_2}{2} (\Phi_2^{\dagger} \Phi_2)^2 + \lambda_3
(\Phi_1^{\dagger} \Phi_1)(\Phi_2^{\dagger} \Phi_2) + \lambda_4
(\Phi_1^{\dagger}
\Phi_2)(\Phi_2^{\dagger} \Phi_1) \nonumber \\
&&+ \left[\frac{\lambda_5}{2} (\Phi_1^{\dagger} \Phi_2)^2 + \rm
h.c.\right] + \left[\lambda_6 (\Phi_1^{\dagger} \Phi_1)
(\Phi_1^{\dagger} \Phi_2) + \rm h.c.\right] \nonumber \\
&& + \left[\lambda_7 (\Phi_2^{\dagger} \Phi_2) (\Phi_1^{\dagger}
\Phi_2) + \rm h.c.\right].
\end{eqnarray}
For the CP-conserving case, all $\lambda_i$ and $m_{12}^2$ are real.
After the electroweak $SU(2)\times U(1)$ symmetry is spontaneously
broken down to $U(1)_{EM}$,
\begin{equation}
\Phi_1=\left(\begin{array}{c} \phi_1^+ \\
\frac{1}{\sqrt{2}}\,(v_1+\phi_1^0+ia_1)
\end{array}\right)\,, \ \ \
\Phi_2=\left(\begin{array}{c} \phi_2^+ \\
\frac{1}{\sqrt{2}}\,(v_2+\phi_2^0+ia_2)
\end{array}\right).
\end{equation}

The mass eigenstates of the five physical scalars can be written as:
\begin{equation}
\left(\begin{array}{c} H\\ h
\end{array}
\right) =\left(
\begin{array}{cc}
\cos\alpha &\sin\alpha\\
-\sin\alpha&\cos\alpha
\end{array}
\right)  \left(
\begin{array}{c}
\phi_1^0\\\phi_2^0
\end{array}
\right),\ \ \
\begin{array}{c}
 A \\H^\pm
 \end{array}
 \begin{array}{l}
 =  -G_1\sin\beta+G_2\cos\beta\\
 =-\phi_1^{\pm}\sin\beta+\phi_2^{\pm} \cos\beta
 \end{array},
 \end{equation}
where $\tan\beta\equiv v_2/v_1$ and $v=\sqrt{v_1^2+v_2^2}\simeq246$
GeV. Their masses are given as [20] \bea
m_A^2&=&\frac{m_{12}^2}{\sinb\cosb}-\frac{v^2}{2}\left(2\lambda_5+\lambda_6\ctb+\lambda_7\tb\right)\nonumber \\
m_{H^\pm}^2&=&m_A^2+\frac{v^2}{2}\left(\lambda_5-\lambda_4\right)\nonumber \\
m_{H,h}^2&=&\frac{1}{2}\left[M_{11}^2+M_{22}^2\pm\sqrt{\left(M_{11}^2-M_{22}^2\right)^2+4\left(M_{12}^2\right)^2}\;\right]
\eea with \begin{equation} M^2=m_A^2\left(\begin{array}{cc}
s^2_\beta & -s_\beta c_\beta \\
-s_\beta c_\beta & c^2_\beta
\end{array}
\right) +v^2 B^2,
\end{equation}
where
\begin{equation}
B^2=\left(\begin{array}{cc} \lambda_1 c^2_\beta+2\lambda_6s_\beta
c_\beta+\lambda_5 s^2_\beta &
\left(\lambda_3+\lambda_4\right)s_\beta c_\beta+\lambda_6 c^2_\beta+\lambda_7 s^2_\beta \\
\left(\lambda_3+\lambda_4\right)s_\beta c_\beta+\lambda_6
c^2_\beta+\lambda_7 s^2_\beta & \lambda_2
s^2_\beta+2\lambda_7s_\beta c_\beta+\lambda_5 c^2_\beta
\end{array}
\right).
\end{equation}
The heavy CP-even Higgs ($H$) and the light CP-even Higgs ($h$) can
be respectively taken as the 125 GeV Higgs. In the physical basis,
$m_h$, $m_H$, $m_A$, $m_{H^\pm}$, $m_{12}^2$, $\sin(\beta-\alpha)$,
$\tan\beta$, $\lambda_6$ and $\lambda_7$ are taken as the free input
parameters. From that point of view, the Higgs masses are
independent on the dimensional constant $m_{12}^2$. The Higgs
spectrum in the minimal supersymmetric standard model (MSSM) can
decouple to the five Higges in 2HDM. To solve some problems such as
unnaturalness of $\mu$ parameter in MSSM, the next-to-minimal
supersymmetric standard model (NMSSM) \cite{nmssm} extends the MSSM
by introducing a gauge singlet superfield $S$ with the
$Z_3$-invariant superpotential given by $W_F +\lambda \hat{H}_u\cdot
\hat{H}_d \hat{S} +\kappa \hat{S}^3/3$. As a result, the NMSSM
predicts one more CP-even Higgs boson and one more CP-odd Higgs
boson in addition to the five Higges. For $\lambda=0$ and
$\kappa=0$, the Higgs spectrum in NMSSM can decouple to MSSM.

In the aligned 2HDM, the two complex scalar doublets couple to the
down-type quarks and charged leptons with aligned Yukawa matrices
\cite{a2h-1,a2h-2}. The Yukawa interactions can be given by \bea -
{\cal L} &=& y_u\,\overline{Q}_L \, \tilde{{ \Phi}}_2 \,u_R +\,y_d\,
\overline{Q}_L\,(\cos{\theta_d}\,{\Phi}_1 \,+\, \sin{\theta_d}\,
{\Phi}_2) \, d_R   \nonumber \\
&&\hspace{-3mm}+ \, y_l\,\overline{l}_L \, (\cos{\theta_l}\,{\Phi}_1
\,+\, \sin{\theta_l}\, {\Phi}_2)\,e_R \,+\, \mbox{h.c.}\,, \eea
where $Q^T=(u_L\,,d_L)$, $L^T=(\nu_L\,,l_L)$, and
$\widetilde\Phi_2=i\tau_2 \Phi_2^*$. $y_u$, $y_d$ and $y_\ell$ are
$3 \times 3$ matrices in family space. $\theta_d$ and $\theta_l$
parameterize the two Higgs doublets couplings to down-type quarks
and charged leptons, respectively. Where a freedom is used to
redefine the two linear combinations of $\Phi_1$ and $\Phi_2$ to
eliminate the coupling of the up-type quarks to $\Phi_1$
\cite{a2h-2}. Table \ref{dlcoup} shows the couplings of two CP-even
Higgs bosons with respect to the SM Higgs boson.

\begin{table}
\caption{The tree-level couplings of the neutral Higgs bosons with
respect to those of the SM Higgs boson. $u$, $d$ and $l$ denote the
up-type quarks, down-type quarks and the charged leptons,
respectively. The angle $\alpha$ parameterizes the mixing of two
CP-even Higgses $h$ and $H$.}
  \setlength{\tabcolsep}{2pt}
  \centering
  \begin{tabular}{|c|c|c|c|c|}
    \hline
     &~~$VV$~$(WW,~ZZ)$~~& ~~~~$u\bar{u}$~~~~ &~~~~ $d\bar{d}$~~~~&~~~~ $l\bar{l}$~~~~\\
    \hline
     $~h~$
     & $\sin(\beta-\alpha)$ & $\frac{\cos\alpha}{\sin\beta}$
     & $-\frac{\sin(\alpha-\theta_d)}{\cos(\beta-\theta_d)}$
     & $-\frac{\sin(\alpha-\theta_l)}{\cos(\beta-\theta_l)}$
     \\
     $~H~$
      & $\cos(\beta-\alpha)$ &$\frac{\sin\alpha}{\sin\beta}$
     &$\frac{\cos(\alpha-\theta_d)}{\cos(\beta-\theta_d)}$
     &$\frac{\cos(\alpha-\theta_l)}{\cos(\beta-\theta_l)}$
     \\
     \hline

      \end{tabular}
\label{dlcoup}
\end{table}

 Now we introduce the renormalizable
Lagrangian of the real single scalar $S$,
\begin{eqnarray}
\mathcal{L}_S&=&-{1\over 2}S^2(\lambda_{1}\Phi_1^\dagger \Phi_1
+\lambda_{2}\Phi_2^\dagger \Phi_2)-{m_{0}^2\over
2}S^2-{\lambda_S\over 4!}S^4\label{potent}.
\end{eqnarray} The linear and cubic terms of the scalar $S$ are
forbidden by a $Z'_2$ symmetry $S\rightarrow -S$. The DM mass and
the interactions with the neutral Higgses are obtained from the Eq.
(\ref{potent}),
\begin{eqnarray}
m_S^2&=&m_0^2+\frac{1}{2}\lambda_1
v^2\cos^2\beta+\frac{1}{2}\lambda_2 v^2\sin^2\beta,\nonumber\\
-\lambda_{h} vS^2h/2&\equiv& -(-\lambda_{1}\sin\alpha\cos\beta+\lambda_{2}\cos\alpha\sin\beta)vS^2h/2,\nonumber\\
-\lambda_{H} vS^2H/2&\equiv&
-(\lambda_{1}\cos\alpha\cos\beta+\lambda_{2}\sin\alpha\sin\beta)vS^2H/2.
\label{dmcoup}\end{eqnarray}

Our previous paper shows detailedly the allowed ranges of $\alpha$,
$\tan\beta$, $\theta_d$, $\theta_l$ and the charged and neutral
Higgses in the aligned 2HDM by the theoretical constraints from
vacuum stability, unitarity and perturbativity as well as the
experimental constraints from the electroweak precision data, flavor
observables and the Higgs searches \cite{leia2h}. In this paper, we
focus on a simplified scenario: (i) The heavy CP-even Higgs $(H)$ is
the discovered 125 GeV Higgs. The masses of pseudoscalar and charged
Higgs are assumed to be heavy enough to avoid the constraints from
the collider experiments and flavor observables. Further, the
electroweak parameter $\rho$ ($\equiv M_W/(M_{Z} cos_{\theta_W})$)
requires their masses to be almost degenerate \cite{leia2h}. (ii)
$\alpha=\beta$ and $\lambda_{H}=0$. As the 125 GeV Higgs, the heavy
CP-even Higgs has the same couplings to the gauge bosons and
fermions as the SM Higgs. $\lambda_{H}=0$ forbids the heavy Higgs
decaying to dark matter. As a result, the heavy Higgs as the 125 GeV
Higgs can fit the Higgs signals well. Only the light CP-even Higgs
$h$ mediates the DM interactions with the SM particles. Its mass is
larger than 62.5 GeV to forbid the decay $H\to hh$. The couplings to
$WW$ and $ZZ$ vanish, and ones to fermions normalized to SM are
$y_u=1/\tan\beta$ for the up-type quarks, $y_d=-tan(\beta-\theta_d)$
for the down-type quarks and $y_l=-tan(\beta-\theta_l)$ for the
charged leptons. In addition, from the Eq. (\ref{dmcoup}), we can
obtain $m_S=m_0$ for $\alpha=\beta$ and $\lambda_{H}=0$.

In our calculations, the involved free parameter of S2HDM+D are
$y_u$ ($\tan\beta$), $y_d$ ($\theta_d$), $y_l$ ($\theta_l$), $m_h$,
$m_S$ and $\lambda_h$. In order to generate the observed spectral
shape of the gamma-ray excess, we fix $m_S=35$ GeV and require
$<\sigma v>_{SS\to b\bar{b}}$ to be in the range of
$1.7-2.3\times10^{-26}cm^3/s$. In 2HDMs, the charged Higgs can give
the additional contributions to the low energy flavor observable
$\Delta m_{B_d}$ and $\Delta m_{B_s}$. The experimental constraints
of $\Delta m_{B_d}$ and $\Delta m_{B_s}$ favor $\tan\beta > 1$ since
the coupling $H^{+}\bar{t}b$ is proportional to $1/\tan\beta$
\cite{leia2h}. In addition, the perturbative of Higgs potential
disfavors the large $\tan\beta$ for the absence of the soft-breaking
term \cite{13050002}. Therefore, we take $0.2<y_u<1.0$
$(1.0<\tan\beta<5.0)$. For such $\tan\beta$ $(\alpha=\beta)$, both
$y_d$ ($-\tan(\beta-\theta_d)$) and $y_l$ ($-\tan(\beta-\theta_l)$)
are allowed to be in the range of $-1.0$ and $0.5$. Here we take
$-1.0 <y_d<-0.2$ which has opposite sign to $y_u$, and favors to
obtain an isospin-violating $S$-nucleon coupling. For simplicity, we
take $y_l=0$ to favor $S$ to annihilate dominantly to $b\bar{b}$.
$\lambda_h$ and $m_h$ are taken to be in the ranges of 0.0001-1.0
and 75-120 GeV, respectively.

\subsection{The spin-independent cross section between $S$ and nucleon}
In this model, the elastic scattering of $S$ on a nucleon receives
the contributions from the $h$ exchange diagrams, which is given as
\cite{sigis},
 \beq \sigma_{p(n)}=\frac{\mu_{p(n)}^{2}}{4\pi m_{S}^{2}}
    \left[f^{p(n)}\right]^{2},
\eeq where $\mu_{p(n)}=\frac{m_Sm_{p(n)}}{m_S+m_{p(n)}}$, \beq
f^{p(n)}=\sum_{q=u,d,s}f_{q}^{p(n)}\mathcal{C}_{S
q}\frac{m_{p(n)}}{m_{q}}+\frac{2}{27}f_{g}^{p(n)}\sum_{q=c,b,t}\mathcal{C}_{S
q}\frac{m_{p(n)}}{m_{q}},\label{fpn} \eeq with $\mathcal{C}_{S
q}=\frac{\lambda_h m_q}{m_h^2}y_q$ . Following the recent study
\cite{1312.4951}, we take
\begin{eqnarray}
f_{u}^{(p)}\approx0.0208,\quad & f_{d}^{(p)}\approx0.0399,\quad &
f_{s}^{(p)}\approx0.0430,
\quad f_{g}^{(p)}\approx0.8963,\nonumber\\
f_{u}^{(n)}\approx0.0188,\quad & f_{d}^{(n)}\approx0.0440,\quad &
f_{s}^{(n)}\approx0.0430,\quad  f_{g}^{(n)}\approx0.8942.
 \label{eq:neuclon-form}
\end{eqnarray}
For the relations $f_{q}^{(p)}=f_{q}^{(n)}$ and
$f_{g}^{(p)}=f_{g}^{(n)}$ are satisfied, the $S$-nucleon coupling is
always isospin-conserved. Conversely, the $S$-nucleon coupling is
violated for the relations are not satisfied, as shown in Eq.
(\ref{eq:neuclon-form}). However, the Higgs couplings to the quarks
($y_d$ and $y_u$) need fine-tuning in order to obtain the large
violating, such as $f^{n}/f^{p}=-0.7$. The recent data on the direct
DM search from LUX put the most stringent constraint on the cross
section \cite{lux}. For the isospin-violating $S$-nucleon coupling,
the scattering rate with the target can be suppressed, thus
weakening the constrains from LUX and XENON100 \cite{XENON100},
especially for $f_n/f_p\simeq-0.7$. Results of direct detection
experiments are often quoted in terms of "normalized-to-nucleon
cross section", which is given by \cite{isospin}
\begin{equation}
\frac{\sigma_p}{\sigma_N^Z} = \frac{\sum_i \eta_i \mu_{A_i}^2 A_i^2}
{\sum_i \eta_i \mu_{A_i}^2 [Z + (A_i - Z) f_n/f_p]^2},
\end{equation}
$\sigma_N^Z$ is the typically-derived DM-nucleon cross section from
scattering off nuclei with atomic number $Z$, assuming isospin
conservation and the isotope abundances found in nature. $\eta_i$ is
the natural abundance of the i-th isotope.

\subsection{Relic density, indirect detection and collider constraints}
In the parameter space taken in the S2HDM+D, the main annihilation
processes include $SS\to q\bar{q}$ and $SS\to gg$ which proceed via
an s-channel $h$ exchange. For the absolute value of $y_d$ is much
less than $y_u$, $SS\to gg$ and $SS\to c\bar{c}$ annihilation
processes can dominate over $SS\to b\bar{b}$. We employ $\textsf{
micrOMEGAs-3.6.9.2}$ \cite{micomega} to calculate the relic density
and the today pair-annihilation cross sections of DM in the inner
galaxy. The Planck collaboration released its relic density as
$\Omega_{c}h^2\pm\sigma=0.1199\pm0.0027$ \cite{planck}, and we
require S2HDM+D to explain the experimental data within $2\sigma$
range.

The heavy CP-even Higgs has the same couplings as SM Higgs, which
can fit the Higgs signals at the LHC well. There is no couplings to
$WW$, $ZZ$ and leptons for the light CP-even Higgs, which favors it
not to be detected at the collider. $\textsf{HiggsBounds-4.1.1}$
\cite{hb-1} is used to implement the exclusion constraints from the
Higgses searches at LEP, Tevatron and LHC at 95\% confidence level.

The ATLAS \cite{mono-atlas} and CMS \cite{mono-cms} collaborations
have published monojet search results, which can be used to place
constraints on the DM-nucleon scattering cross section. For the
scalar portal DM, the DM interactions with the light quarks are
proportional to quark mass, leading to suppressing the
monojet+$\Eslash_T$ signal sizably. Therefore, the current monojet
searches for DM at the LHC appears to provide no stronger
constraints on the S2HDM+D than the direct detection from the LUX
experiment \cite{mono-atlas,mono-cms,mono}.

\section{results and discussions}
\begin{figure}[tb]
 \epsfig{file=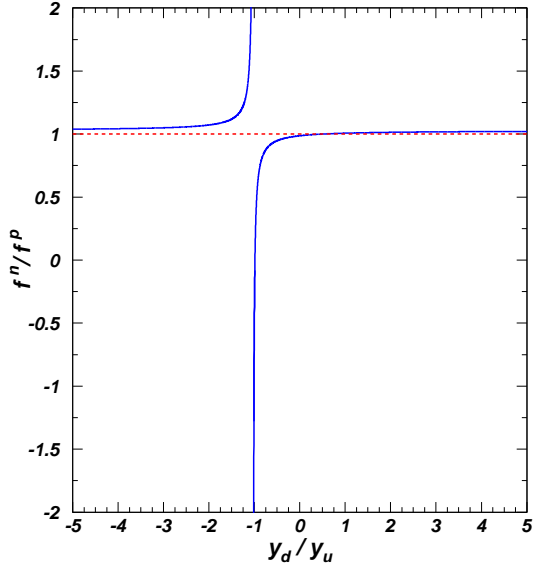,height=7.5cm}
\vspace{-0.4cm} \caption{$f^n/f^p$ versus $y_d/y_u$.} \label{fnfp}
\end{figure}

Since the hadronic quantities in the spin independent $S$-nucleon
scattering are fixed, $f^n/f^p$ only depends on the normalized
factors of Yukawa couplings, $y_u$ and $y_d$. Fig. \ref{fnfp} shows
$f^n/f^p$ versus $y_d/y_u$. We find that $f^n/f^p$ is very sensitive
to $y_d/y_u$ for $y_d/y_u$ is around -1.0, and very close to 1.0 for
$y_d/y_u>0$. In the following discussions, we will focus on the
surviving samples with $-1.0< f^n/f^p<1.0$ where the constraint from
the LUX experiment can be weakened. $f^n/f^p$ in such range favors
$y_d/y_u<$ 0, which is the reason why we choose $y_d$ to have
opposite sign to $y_u$.

\begin{figure}[tb]
 \epsfig{file=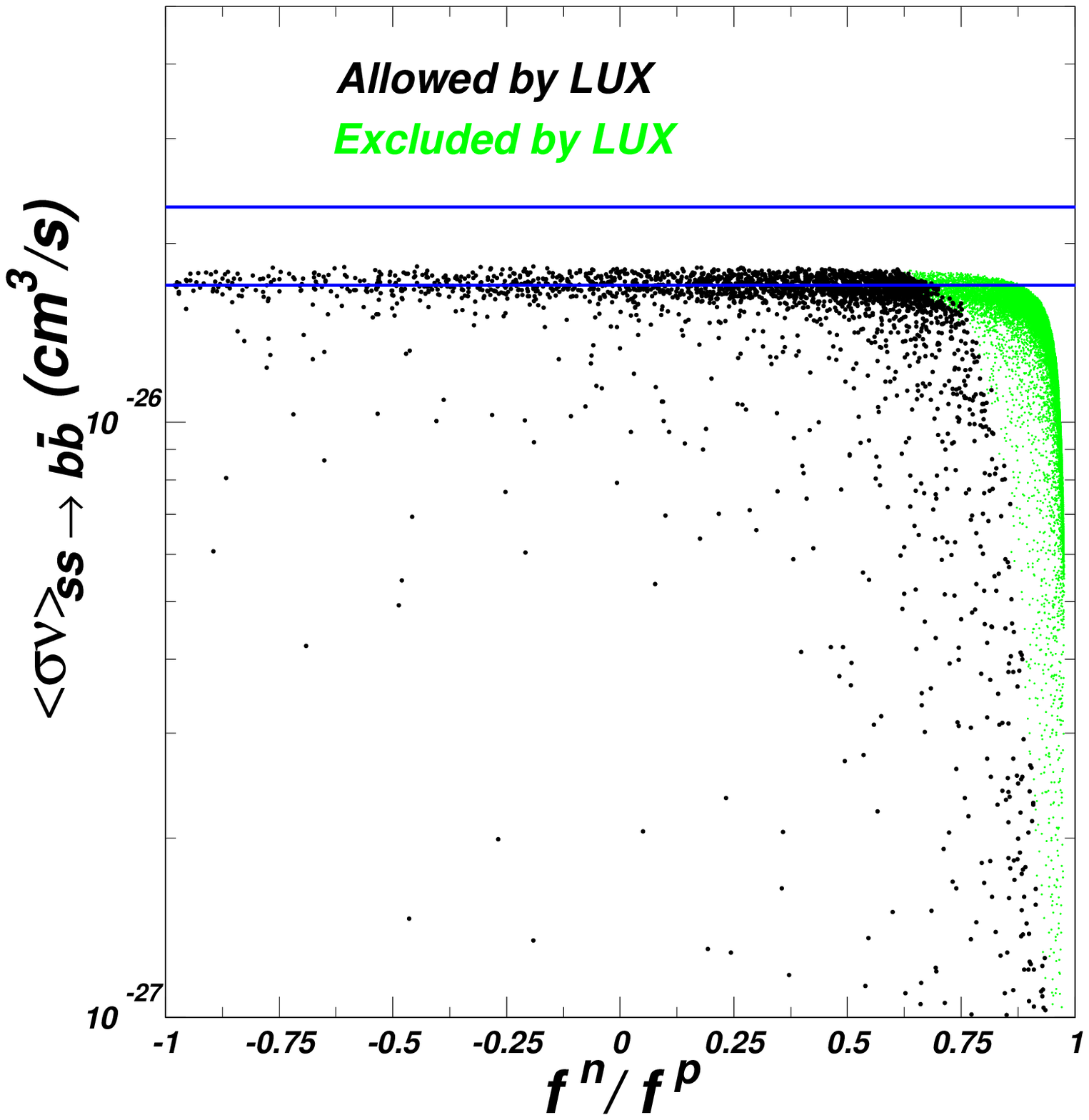,height=7.0cm}
  \epsfig{file=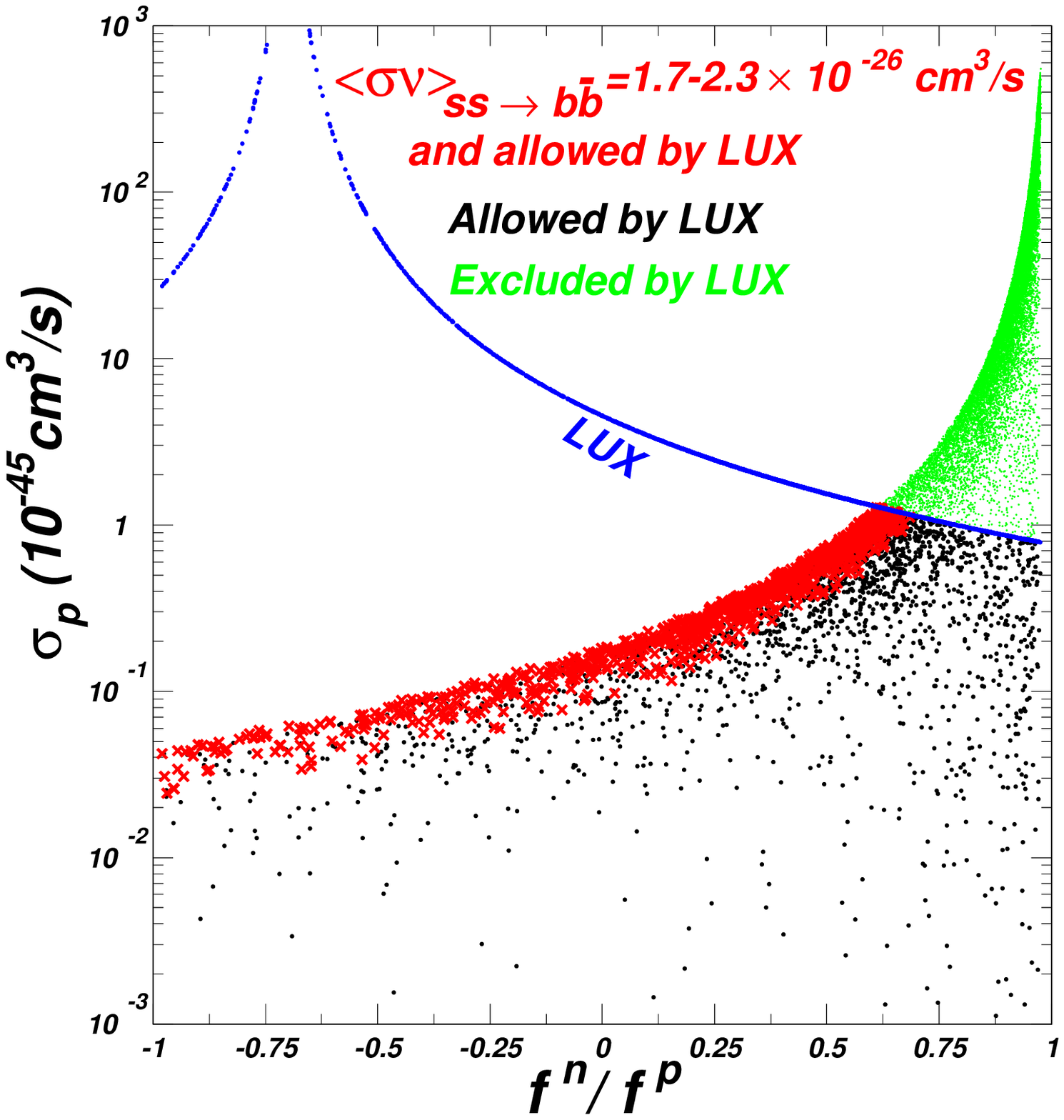,height=7.0cm}
\vspace{-0.4cm} \caption{The scatter plots of surviving samples
projected on the planes of  $<\sigma v>_{SS\to b\bar{b}}$ versus
$f^n/f^p$ and $\sigma_p$ versus $f^n/f^p$. The two horizontal lines
in the left panel denote $<\sigma v>_{SS\to
b\bar{b}}=1.7\times10^{-26}cm^{3}/s$ and
$2.3\times10^{-26}cm^{3}/s$.} \label{sigbb-si}
\end{figure}

In Fig. \ref{sigbb-si}, we project the surviving samples on the
planes of $<\sigma v>_{SS\to b\bar{b}}$ versus $f^n/f^p$ and
$\sigma_p$ versus $f^n/f^p$, respectively. The left panel shows
that, for $-1< f^n/f^p<0.7$, $<\sigma v>_{SS\to b\bar{b}}$ can be in
the range of $1.7-2.3\times10^{-26}cm^3/s$ while $\sigma_p$ is below
the upper bound from the LUX experiment. For $f^n/f^p$ is very close
to 1.0, $<\sigma v>_{SS\to b\bar{b}}$ as low as $10^{-27}cm^3/s$ is
still not allowed by the LUX constraint. The right panel shows that
the maximal value of $\sigma_p$ decreases as $f^n/f^p$ varies from
1.0 to -1.0, and $\sigma_p$ is smaller than the upper bound of LUX
by several orders of magnitude for $f^n/f^p$ is around -0.7.

\begin{figure}[tb]
 \epsfig{file=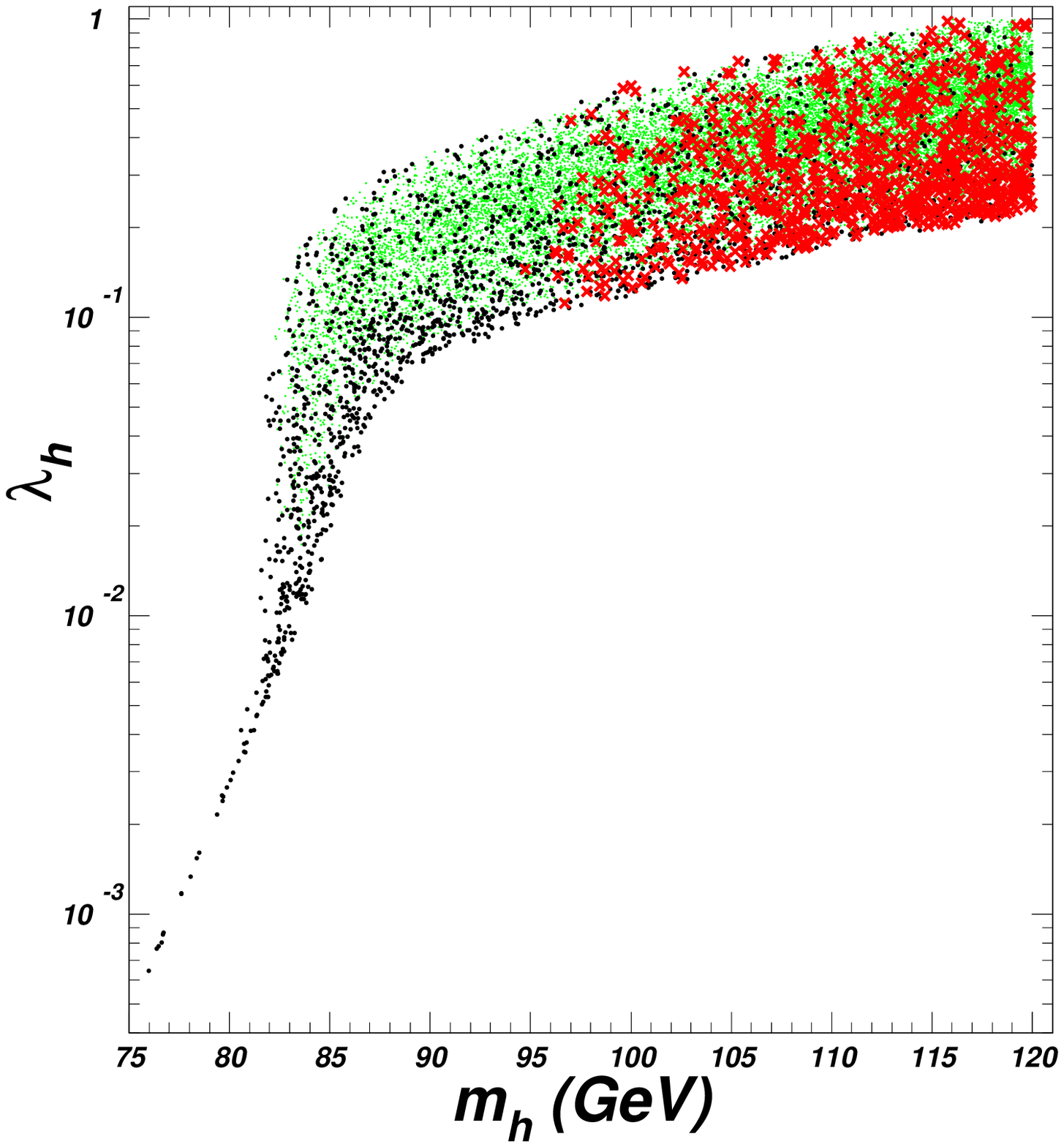,height=5.8cm}
  \epsfig{file=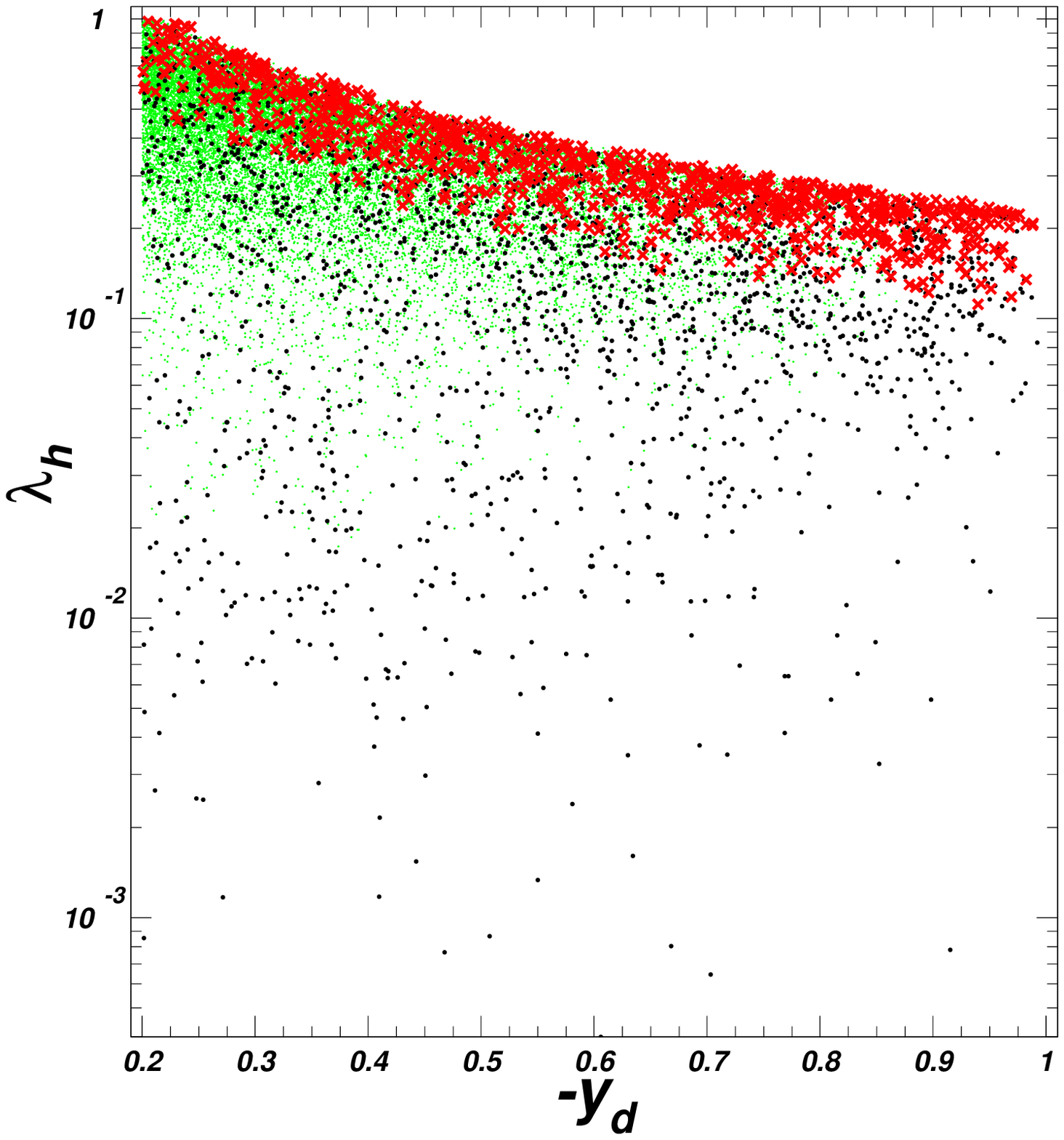,height=5.8cm}
     \epsfig{file=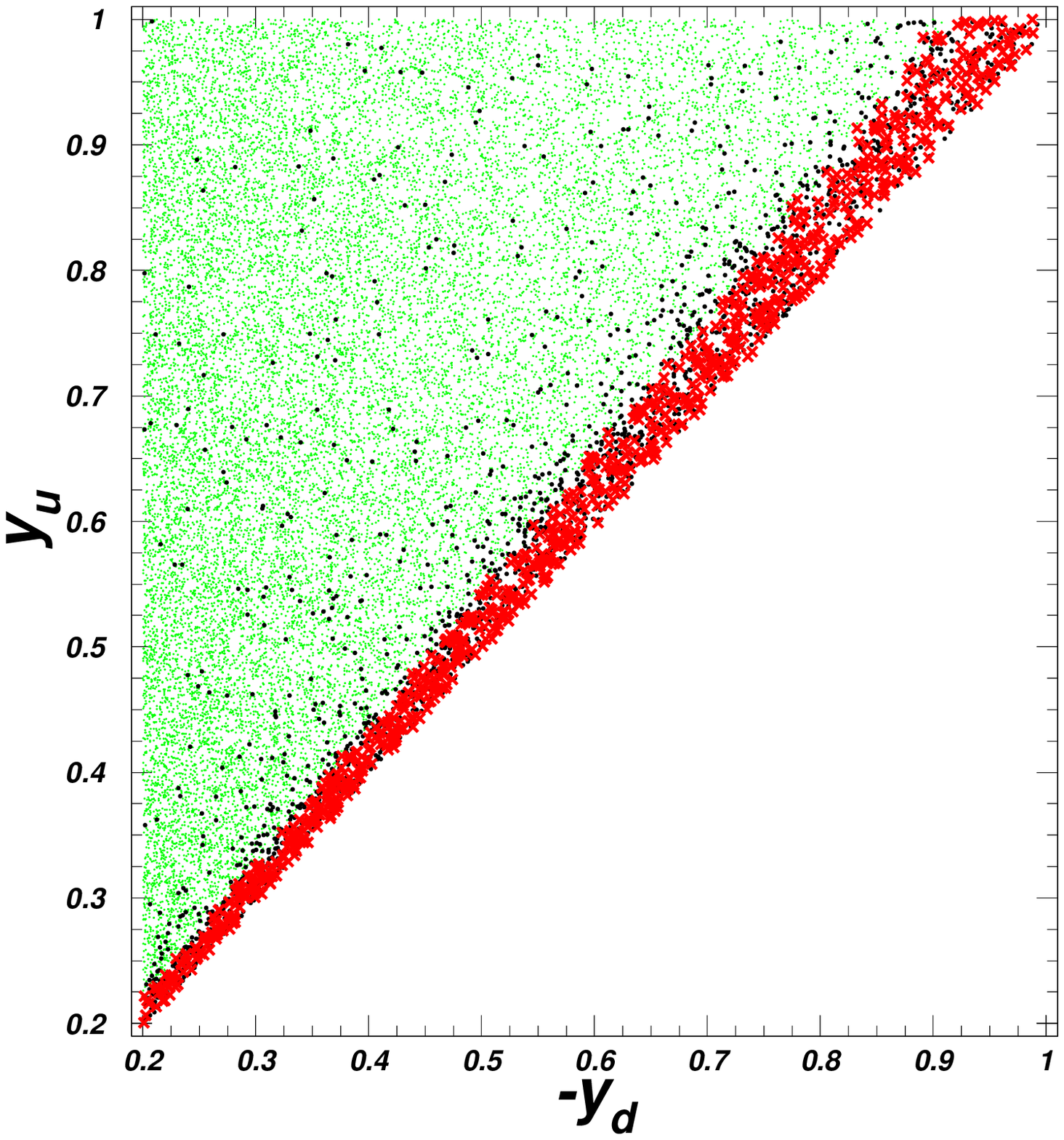,height=5.79cm}
\vspace{-1.2cm} \caption{Same as Fig. \ref{sigbb-si}, but projected
on the planes of $\lambda_h$ versus $m_h$, $\lambda_h$ versus
$-y_d$, and $y_u$ versus $-y_d$, respectively.} \label{lam}
\end{figure}

In Fig. \ref{lam}, we project the surviving samples on the planes of
$\lambda_h$ versus $m_h$, $\lambda_h$ versus $-y_d$, and $y_u$
versus $-y_d$, respectively. For the surviving samples which can
explain the gamma-ray excess validly: The middle panel shows the
lower bound of $\lambda_h$ is 0.1 for $-y_d=1.0$, and enhanced to
0.6 as $-y_d$ decreases to 0.2. The left panel shows the lower bound
of $\lambda_h$ is visibly enhanced for the large $m_h$, such as
$m_h=$ 120 GeV. For $m_h/2$ approaches to $m_S$ (35 GeV),
$\lambda_h$ can be much smaller than 0.1 to achieve the correct
relic abundance since the integral in the calculation of thermal
average can be dominated by the resonance at $s = m_h^2$ even if
$m_S$ is below $m_h/2$. However, such small $\lambda_h$ will
suppress sizably the scattering of DM off nuclei and even the today
pair-annihilation of DM into $b\bar{b}$ which leads to fail to
explain the gamma-ray excess. The right panel shows that $y_d/y_u$
is required to be around -1.0 where $f^n/f^p$ is in the range of
-1.0 and 0.7 (see Fig. \ref{fnfp} and Fig. \ref{sigbb-si}), and DM
annihilates dominantly into $b\bar{b}$.

\section{Conclusion}
In this note, we add a scalar DM to the aligned 2HDM, and focus on a
simplified scenario, which very economically implements the
following two characteristics: (i) The heavy CP-even Higgs is the
discovered 125 GeV Higgs boson, which has the same couplings to the
gauge bosons and fermions as the SM Higgs. (ii) Only the light
CP-even Higgs mediates the DM interactions with SM particles, which
has no couplings to $WW$ and $ZZ$, but the independent couplings to
the up-type quarks, down-type quarks and charged leptons. We find
that the tension between $<\sigma v>_{SS\to b\bar{b}}$ and the
constraint from LUX induced by the scalar portal DM can go away for
the isospin-violating $S$-nucleon coupling, $-1.0< f^n/f^p<0.7$.
Being consistent with the constraints from the relic density of
Planck, the direct detection of LUX, the Higgs searches at the
collider and the other relevant experiments, the model can give a
valid explanation for the galactic center gamma-ray excess in the
proper ranges of $\lambda_h$, $m_h$, $y_u$ and $y_d$.

\section*{Acknowledgment}
This work was supported by the National Natural Science Foundation
of China (NNSFC) under grant No. 11105116.


\begin{thebibliography}{99}
\bibitem{10gev} L. Goodenough and D. Hooper, arXiv:0910.2998; D.
Hooper and L. Goodenough, \PLB697, 412 (2011); D. Hooper and T.
Linden, \PRD84, 123005 (2011); K. N. Abazajian and M. Kaplinghat,
\PRD86, 083511 (2012); D. Hooper, C. Kelso, F. S. Queiroz,
arXiv:1209.3015; D. Hooper and T. R. Slatyer, Phys. Dark Univ.
{\bf2}, 118 (2013); C. Gordon and O. Macias, \PRD88, 083521 (2013);
W. -C. Huang, A. Urbano and W. Xue, arXiv:1307.6862.

\bibitem{hoopbb} T. Daylan, D. P. Finkbeiner, D. Hooper, T. Linden, S. K. N.
Portillo, N. L. Rodd and T. R. Slatyer, arXiv:1402.6703.


\bibitem{lux} D. S. Akerib et al. [LUX Collaboration], \PRL112, 091303
(2014).

\bibitem{1404.3716} C. Boehm, M. J. Dolan, C. McCabe, M. Spannowsky, C. J.
Wallace, arXiv:1401.6458; S. Ipek, D. McKeen, A. E. Nelson,
arXiv:1404.3716.


\bibitem{flavorb} P. Agrawal, B. Batell, D. Hooper and T. Lin,
arXiv:1404.1373.

\bibitem{1408.2106-69} S. Chatrchyan et al. [CMS Collaboration], \EPJC74, 2980 (2014).

\bibitem{1408.2106-56} G. Belanger, B. Dumont, U. Ellwanger, J. Gunion, and S. Kraml,
                       \PRD88, 075008 (2013).

\bibitem{1408.2106} A. Drozd1, B. Grzadkowski1, J. F. Gunion, Y.
                    Jiang, arXiv:1408.2106.


\bibitem{gamtautau} T. Lacroix, C. Boehm, J. Silk, arXiv:1403.1987;
K. N. Abazajian, N. Canac, S. Horiuchi, M. Kaplinghat,
arXiv:1402.4090.

\bibitem{onshell} C. Boehm, M. J. Dolan, C. McCabe, arXiv:1404.4977;
M. Abdullah, A. DiFranzo, A. Rajaraman, T. M. P. Tait, P. Tanedo, A.
M. Wijangco, arXiv:1404.6528; A. Martin, J. Shelton, J. Unwin,
arXiv:1405.0272; P. Ko, W.-I. Park, Y. Tang, arXiv:1404.5257.

\bibitem{gambbother} Z. kang, P. Ko, T. Li, Y. Liu, arXiv:1403.7742;
T. Lacroix, C. Boehm, J. Silk, arXiv:1403.1987; A. Hektor, L.
Marzola, arXiv:1403.3401; A. Alves, S. Profumo, F. S. Queiroz, W.
Shepherd, arXiv:1403.5027; A. Berlin, D. Hooper, S. D. McDermott,
arXiv:1404.0022; E. Izaguirre, G. Krnjaic, B. Shuve,
arXiv:1404.2018; Q. Yuan, B. Zhang, arXiv:1404.2318; D. G. Cerdeno,
M. Peiro, S. Robles, arXiv:1404.2572; D. K. Ghosh, S. Mondal, I.
Saha, arXiv:1405.0206; A. Berlin, P. Gratia, D. Hooper, S. D.
McDermott, arXiv:1405.5204;
 T. Basak, T. Mondal, arXiv:1405.4877;
 K. Agashe, Y. Cui, L. Necib, J. Thaler, arXiv:1405.7370;
J. M. Cline, G. Dupuis, Z. Liu, W. Xue, arXiv:1405.7691; N. Okada,
O. Seto, \PRD89, 043525 (2014); M. S. Boucenna, S. Profumo, \PRD84,
 055011 (2011); K. Kong, J.-C. Park, arXiv:1404.3741; B. Kyae, J.-C. Park,
arXiv:1310.2284; K. P. Modak, D. Majumdar, S. Rakshit,
arXiv:1312.7488; E. Hardy, R. Lasenby, J. Unwin, arXiv:1402.4500.


\bibitem{a2h-1} A. Pich, P. Tuzon, \PRD80, 091702 (2009).
\bibitem{a2h-2} V. Barger, L. L. Everett, H. E. Logan and G. Shaughnessy, \PRD88, 115003 (2013).

\bibitem{125higgs} S. Chatrchyan et al. [CMS Collaboration], \PLB716, 30
(2012); G. Aad et al. [ATLAS Collaboration], \PLB716, 1 (2012).

\bibitem{ii2hdm-dm} Y. Cai, T. Li, \PRD88, 115004 (2013).

\bibitem{iii2hdm-dm} X.-G. He, J. Tandean, \PRD88, 013020 (2013).

\bibitem{htm-dm} L. Wang, X.-F. Han, \PRD87, 015015 (2013).

\bibitem{130ray} C. Weniger, \JCAP1208, 007 (2012).

\bibitem{2h-poten} R. A. Battye, G. D. Brawn, A. Pilaftsis, \JHEP1108, 020 (2011).

\bibitem{higgmass} D. Eriksson, J. Rathsman, O. Stal, \CPC181, 189-205 (2010); J. F.
Gunion and H. E. Haber, \PRD67, 075019 (2003).

\bibitem{nmssm} U. Ellwanger, C. Hugonie and A. M. Teixeira, \PR496, 1 (2010).

\bibitem{leia2h} L. Wang, X.-F. Han, \JHEP04, 128 (2014).

\bibitem{13050002} B. Coleppa, F. Kling, S. Su, \JHEP01, 161 (2014).

\bibitem{sigis} G. Jungman, M. Kamionkowski, K. Griest, \PR267, 195
(1996); M. A. Shifman, A. I. Vainshtein and V. I. Zakharov, \PLB78,
443 (1978).

\bibitem{1312.4951} A. Crivellin, M. Hoferichter, M. Procura, \PRD89, 054021
(2014).

\bibitem{XENON100}  E. Aprile et al. [XENON100 Collaboration], \PRL109, 181301 (2012).

\bibitem{isospin} J. L. Feng, J. Kumar, D. Marfatia, D. Sanford, \PLB703, 124-127
(2011).

\bibitem{micomega} G. Belanger, F. Boudjema, A. Pukhov, A. Semenov,
arXiv:1305.0237.

\bibitem{planck} P. Ade et al. [Planck Collaboration], arXiv:1303.5076.

\bibitem{hb-1} P. Bechtle, O. Brein, S. Heinemeyer, G. Weiglein, K. E.
Williams, \CPC181, 138-167 (2010); P. Bechtle, O. Brein, S.
Heinemeyer, O. St{\aa}l, T. Stefaniak, G. Weiglein, K. E. Williams,
\EPJC74, 2693 (2014).

\bibitem{mono-atlas} G. Aad et al. [ATLAS Collaboration], \JHEP1304, 075
(2013).
\bibitem{mono-cms} S. Chatrchyan et al. [CMS
Collaboration], \JHEP1209, 094 (2012).

\bibitem{mono} N. Zhou, D. Berge, L. Wang, D. Whiteson and T. Tait,
arXiv:1307.5327; T. Lin, E. W. Kolb and L. -T. Wang, \PRD88, 063510
(2013).

\end{thebibliography}
\end{document}